\documentclass[aps,twocolumn,prb,superscriptaddress,amsmath,amssymb]{revtex4-1}

\usepackage{amsmath}
\usepackage{amsfonts}
\usepackage{amssymb}
\usepackage{dcolumn}
\usepackage{bm}
\usepackage{txfonts}
\usepackage{ulem}
\usepackage{braket}
\usepackage{here}
\usepackage[dvipdfmx]{graphicx,hyperref}
\usepackage[usenames]{xcolor}

\def\journal #1#2#3#4{{#1} {\bf #2}, #3 (#4).}

\begin{document}
\title{
Multiple Dirac Cones and Topological Magnetism in Honeycomb-Monolayer Transition Metal Trichalcogenides
}

\author{Yusuke Sugita}
\affiliation{Department of Applied Physics, University of Tokyo, Bunkyo, Tokyo 113-8656, Japan}
\author{Takashi Miyake}
\affiliation{CD-FMat, National Institute of Advanced Industrial Science and Technology (AIST), Tsukuba, Ibaraki 305-8568, Japan}
\author{Yukitoshi Motome}
\affiliation{Department of Applied Physics, University of Tokyo, Bunkyo, Tokyo 113-8656, Japan}

\date{\today}
\begin{abstract}
The discovery of monolayer graphene has initiated two fertile fields in modern condensed matter physics, Dirac semimetals and atomically-thin layered materials.
When these trends meet again in transition metal compounds, which possess spin and orbital degrees of freedom and strong electron correlations, more exotic phenomena are expected to emerge in the cross section of topological states of matter and Mott physics.
Here, we show by using {\it ab initio} calculations that a monolayer form of transition metal trichalcogenides (TMTs), which has a honeycomb network of $4d$ and $5d$ transition metal cations, may exhibit multiple Dirac cones in the electronic structure of the half-filled $e_g$ orbitals.
The Dirac cones are gapped by the spin-orbit coupling under the trigonal lattice distortion, and hence, can be tuned by tensile strain. 
Furthermore, we show that electron correlations and carrier doping turn the multiple-Dirac semimetal into a topological ferromagnet with high Chern number.
Our findings raise the honeycomb-monolayer TMTs to a new paradigm to explore correlated Dirac electrons and topologically-nontrivial magnetism.
\end{abstract}

\maketitle

%%%%% Introduction %%%%%
\section{Introduction}
Since the success of exfoliation of a monolayer graphene~\cite{Novoselov666}, atomically-thin layered materials have grown as one of the leading themes in modern condensed matter physics.
In particular, van der Waals (vdW) materials, composed of atomic layers bounded via weak vdW forces, have received great attention. 
Electrons confined in an atomically thin layer exhibit drastically distinct behavior from the bulk form. 
The archetypal example is the Dirac electrons in a monolayer graphene, which show anomalous transport behavior, e.g., the anomalous integer quantum Hall effect~\cite{novoselov2005two,zhang2005experimental} and the Klein tunneling~\cite{katsnelson2006chiral,young2009quantum}.
Another example is the valley degree of freedom in the monolayer form of transition metal dichalcogenides~\cite{doi:10.1021/nl903868w}, which has been intensively studied toward valleytoronics devices~\cite{wang2012electronics,xu2014spin}.
Furthermore, heterostructures of different vdW materials have provided a new platform for novel functionalities never seen in bulk compounds~\cite{geim2013van,novoselov20162d}. 

Through the intensive research in the past decade, a lot of efforts have been made to find atomically-thin {\it magnetic} materials.
Among many candidates, a family of transition metal trichalcogenides (TMTs) has gained increasing interests, both from theoretical proposals of monolayer magnetism~\cite{PhysRevB.91.235425,PhysRevB.94.184428} and experimental reports on the mono and few-layer forms~\cite{doi:10.1021/acsnano.5b05927,C5TC03463A,kuo2016exfoliation,doi:10.1021/acs.nanolett.6b03052,2053-1583-3-3-031009,Gong2017}.
In addition, not only the magnetism but also anomalous electronic and transport properties are predicted in the presence of the relativistic spin-orbit coupling (SOC), e.g., the spin-valley coupling~\cite{Li05032013}, the magnon spin Nernst effect~\cite{PhysRevLett.117.217202}, and the gate-controllable magneto-optic Kerr effect~\cite{PhysRevLett.117.267203}. 
Thus, the atomically-thin layered TMTs are expected to provide a unique cross section between strong electron correlations and the SOC, but their potential remains unexplored.

In this paper, we theoretically propose that monolayer TMTs with a honeycomb network of $4d$ and $5d$ transition metals would host a new playground for correlated Dirac electrons and topologically-nontrivial magnetism.
By {\it ab initio} calculations, we show that the TMTs with group 10 transition elements have semimetallic band structures with multiple Dirac cones in the half-filled $e_g$ orbitals in the paramagnetic state.
We find that the multiple Dirac cones originate in electron transfers on a hidden honeycomb superstructure emergent from spatially anisotropic $d$ orbitals with large hybridization with the neighboring ligand $p$ orbitals. 
We also show that the SOC gaps out these Dirac cones in the presence of the trigonal lattice distortions, and hence, the mass gap can be flexibly tuned by the tensile strain.
In addition, by the mean-field analysis for an effective multi-orbital Hubbard model, we elucidate that electron correlations and chemical doping potentially change the multiple-Dirac semimetals into a topological ferromagnet with high Chern numbers. 

This paper is organized as follows.
In Sec.~\ref{cry}, we introduce the method of {\it ab initio} calculations used in this paper.
In Sec.~\ref{res}, we show the results of the {\it ab initio} calculations and the mean-field analysis of an effective multi-orbital Hubbard model for monolayer TMT.
Section~\ref{sum} is devoted to the summary and the discussion of future issues.
In Appendix A, we show the details of optimized crystalline structures, {\it ab initio} band structures, and transfer integrals of monolayer TMTs.
In Appendix B, we provide the information on distortions under tensile strain.
The {\it ab initio} results for magnetism of monolayer TMTs are presented in Appendix C.

%%%%% Crystal and ab initio calc. %%%%%
\section{
{\it ab initio}  calculation
}
\label{cry}
\begin{figure*}[t]
\centering
\includegraphics[width=2.0\columnwidth]{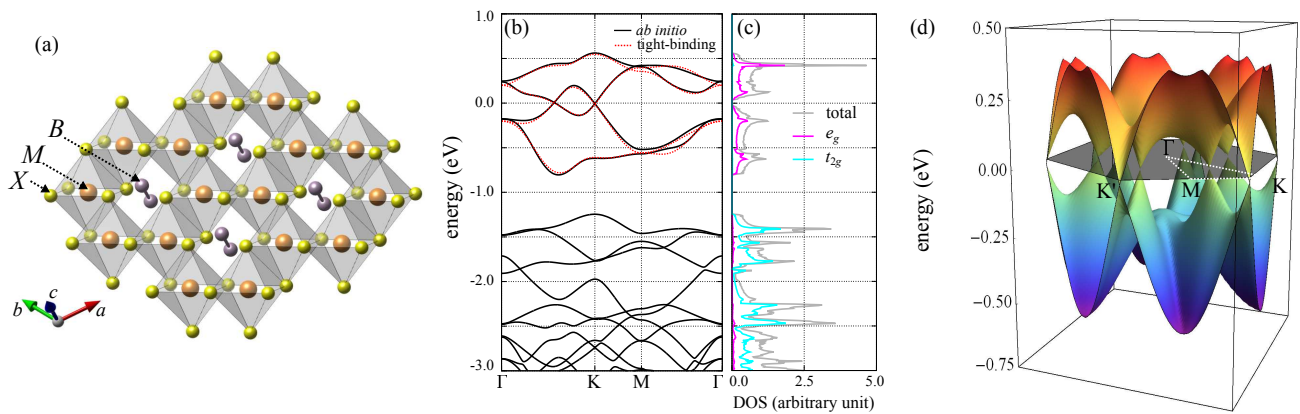}
\caption{
(a)~Schematic picture of a honeycomb-monolayer TMT, whose chemical formula is given as $MBX_3$.
The orange, purple, and yellow spheres denote the transition metals $M$, $B$, and chalcogens $X$, respectively.
The gray octahedra indicate the edge-sharing $MX_6$.
$M$ forms a honeycomb network, while $B$ composes a dimer located at the center of each hexagon of the honeycomb structure. 
(b)~Electronic band structure of a monolayer PdPS$_{3}$ in the paramagnetic state without the SOC.
The Fermi level is set to zero.
The black solid lines represent the band dispersions obtained by {\it ab initio} calculations, while the red dotted ones are those by the tight-binding model for the $e_g$ bands with the transfer integrals between MLWFs up to fifth neighboring Pd cations (see Table~\ref{hwr}). 
(c)~Total density of states (DOS) and projected DOS for the Pd $d$ orbitals.
(d)~3D plot of the two bands near the Fermi level.
The multiple Dirac nodes are formed at the K and K' points and around the midpoints in the $\Gamma$-K lines in the first Brillouin zone indicated by the gray hexagon.
}
\label{setup}
\end{figure*}

The chemical formula for TMTs is generally given by $MBX_3$, where $M$ is transition metals, $B$=P, Si, or Ge, and $X$ is chalcogens.
TMTs have vdW layered structures, whose stacking manner depends on the compounds~\cite{ZAAC:ZAAC19733960305,BREC19863,OUVRARD198827,carteaux1995crystallographic}. 
In each layer, transition metal cations $M$ comprise a honeycomb network by sharing the edges of $MX_6$ octahedra, and $B_2$ dimers locate at the centers of the hexagons of the honeycomb network [Fig.~\ref{setup}(a)]. 

In this paper, we focus on a monolayer form of TMTs with $B$=P and $X$=S and Se.
In this case, the nominal valence of the transition metal cation is $M^{2+}$. 
We note that all 3$d$, 4$d$, and 5$d$ transition metal elements belonging to group 10 and 12 can take the stable divalent oxidation state~\cite{tagkey1997ii}.
Indeed, $M$P$X_{3}$ with $M$=Ni, Pd, Zn, Cd, and Hg have been synthesized~\cite{ZAAC:ZAAC19733960305}.
In the following, we consider group 10 elements, $M$=Ni, Pd, and Pt.

We calculate the electronic band structures of monolayer $M$P$X_3$ by {\it ab initio} calculations based on the generalized gradient approximation (GGA).
In the calculations, we used OpenMX code~\cite{openmx}, which is based on a linear combination of pseudoatomic orbital formalism~\cite{PhysRevB.67.155108,PhysRevB.69.195113}.
We adopted the Perdew-Burke-Ernzerhof GGA functional in density functional theory~\cite{PhysRevLett.77.3865} and a $30\times30\times1$ $\bm{k}$-point mesh for the calculations of the self-consistent electron density and the structure relaxation.
We inserted vacuum space greater than 10\AA~between monolayers and fully relaxed the primitive vectors and atomic positions in the unit cell with the convergence criterion 0.01 eV/\AA~about the inter-atomic forces.

For all combinations of $M$=Ni, Pd, and Pt and $X$=S and Se, we performed full structural optimization in the paramagnetic state without the SOC, starting from the reported crystalline data of NiPS$_3$ or PdPS$_3$~\cite{ZAAC:ZAAC19733960305}.
We confirmed that all the cases stably converge on the similar structure, as shown in Appendix~\ref{app1}.
The SOC is incorporated by the relativistic {\it ab initio} calculations for the optimized structures. 
In the following sections, we will discuss the material trend of $M$P$X_3$ on the basis of PdPS$_3$, as it represents the typical band structure of $M$P$X_3$ and locates between weakly correlated $5d$ and strongly correlated $3d$ systems. 
(Indeed, we will discuss that the compound might be close to the border between paramagnetic and antiferromagnetic phases.)
For the microscopic analyses based on the tight-binding models, we construct the maximally localized Wannier functions (MLWFs)~\cite{PhysRevB.56.12847,PhysRevB.65.035109} for the $e_{g}$ bands and evaluate the transfer integrals between them via a code implemented in OpenMX~\cite{openmx}. 
We also compute the magnetic solutions by using GGA scheme without the SOC under the full structural optimization, as shown in Appendix~\ref{app3}.

%%%%% Results %%%%%
\section{Results}
\label{res}

%%%%% Multiple Dirac cones in honeycomb-monolayer TMTs %%%%%
\subsection{Multiple Dirac cones}
\label{sec:multipleDirac}

We show the electronic band structures in the paramagnetic state obtained by {\it ab initio} calculations without the SOC.
Figures \ref{setup}(b) and \ref{setup}(c) show the representative results for PdPS$_{3}$.
The Pd $d$-orbital levels are split into two groups, $e_{g}$ and $t_{2g}$, due to the crystalline electric fields of octahedral ligands.
As Pd$^{2+}$ is in the $d^{8}$ electron configuration, the lower-energy $t_{2g}$ manifold is fully occupied and the higher-energy $e_{g}$ manifold is half-filled. 
Remarkably, the $e_g$ bands have two crossing points at the K point and around the midpoint on the $\Gamma$-K line in the Brillouin zone [Fig.~\ref{setup}(b)], and the projected density of states are almost zero at the Fermi level [Fig.~\ref{setup}(c)].
We find that the crossings are the Dirac cones, as shown in Fig.~\ref{setup}(d): two electronic bands near the Fermi level give rise to eight Dirac cones (two on the zone boundary, K and K',  and other six inside).
We confirm that the multiple Dirac cones are shared by other monolayer TMTs with $M$=Ni, Pd, and Pt (see Appendix~\ref{app1} for details).

\begin{figure}[t]
\centering
\includegraphics[width=1.0\columnwidth]{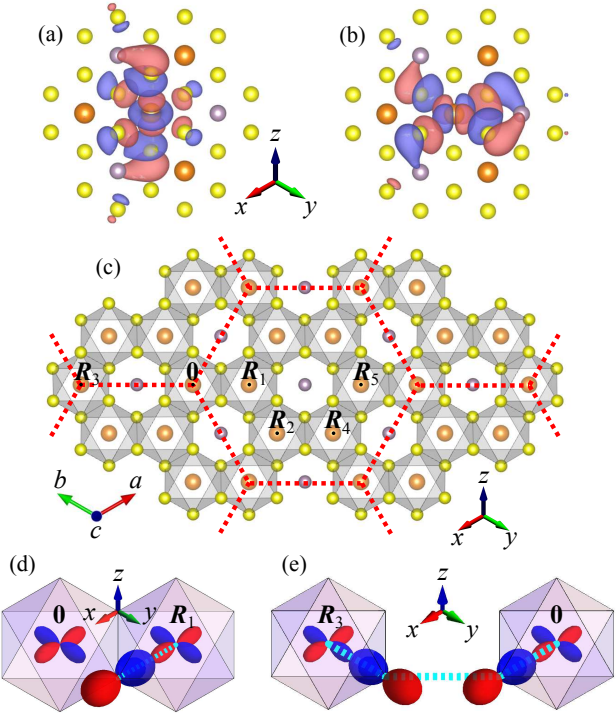}
\caption{
(a) and (b)~Contour-surface plot of MLWFs, which are obtained from two initial states, $d_{3z^{2}-r^{2}}$ and $d_{x^{2}-y^{2}}$ orbitals, respectively.
The red (blue) surfaces indicate the positive (negative) isosurface at +0.03 (-0.03). 
Both wave functions are not localized around the Pd site but fairly extended to neighboring S sites. 
(c)~Atomic positions used in the calculation of transfer integrals summarized in Table~\ref{hwr}.
$\bm{R}_i$ represents the $i$th neighbor site to the 0 site.
The red dotted lines indicate a honeycomb superstructure composed of the third neighbor bonds. 
(d) and (e)~Schematic pictures of first and third neighbor hopping processes via the ligand $p$ orbitals.
}
\label{mlwf}
\end{figure}

In order to clarify the microscopic origin of the multiple Dirac cones, we show the MLWFs obtained from two initial states, $d_{3z^{2}-r^{2}}$ and $d_{x^{2}-y^{2}}$, in Figs.~\ref{mlwf}(a) and \ref{mlwf}(b), respectively. 
Both MLWFs well extend over the neighboring S sites, indicating the importance of indirect hopping processes via the ligand $p$ orbitals. 
Table~\ref{hwr} shows the representative transfer integrals between the two types of MLWFs for the Pd-Pd bonds up to fifth neighbors [see Fig.~\ref{mlwf}(c)].
(See Appendix~\ref{app1} for the extended list of transfer integrals.)
We construct a tight-binding model by using these transfer integrals and confirm that the model well reproduces the {\it ab initio} band structure [see Fig.~\ref{setup}(b)].
Interestingly, the most dominant electron transfer is not for nearest neighbors but the third neighbors.
This is understood from the fact that the indirect $d$-$p$-$d$ hoppings between nearest neighbors are almost forbidden, while the $d$-$p$-$p$-$d$ ones for third neighbors are substantial, as shown in Figs.~\ref{mlwf}(d) and \ref{mlwf}(e).
We note that a similar argument was made for magnetic exchange interactions~\cite{PhysRevB.91.235425}.

\begin{table}[b]
\begin{ruledtabular}
\begin{tabular}{l|rrrrrr}
$(m,n)$	&$\bm{R}_{1}$ &$\bm{R}_{2}$ & $\bm{R}_{3}$ & $\bm{R}_{4}$ & $\bm{R}_{5}$\\
\hline
$(3z^{2}-r^{2},3z^{2}-r^{2})$	&$-87$	&$-9$	&$-38$	&$4$		&$-12$\\
$(3z^{2}-r^{2},x^{2}-y^{2})$	&$0$		&$22$	&$0$		&$-8$	&$0$\\
$(x^{2}-y^{2},3z^{2}-r^{2})$	&$0$		&$18$	&$0$		&$-8$	&$0$\\
$(x^{2}-y^{2},x^{2}-y^{2})$		&$-70$	&$14$	&$304$	&$7$		&$30$\\
\end{tabular}
\end{ruledtabular}
\caption{
Transfer integrals between MLWFs.
Each value in the table means $\bra{m,\bm{0}} H \ket{n,\bm{r}}$, where $H$ is the Hamiltonian of the system and $\ket{m,\bm{r}}$ is the $d_{m}$-like MLWF at site $\bm{r}$ ($m$ = $3z^{2}-r^{2}$ or $x^{2}-y^{2}$).
We take $\bm{r}=\bm{R}_{i}$ ($i$ = 1, 2, 3, 4, or 5) illustrated in Fig.~\ref{mlwf}(c).
The unit of transfer integrals is in meV.
}
\label{hwr}
\end{table}

The dominant third neighbor transfers explain the origin of the multiple Dirac cones. 
As well-known in graphene, the nearest neighbor transfers produce the Dirac cones at the zone corners, the K and K' points.
This is also the case for the $e_{g}$ electron systems~\cite{xiao2011interface}. 
On the other hand, the third neighbor transfers bring about Dirac cones at the additional six points inside the first Brillouin zone: the network of the third neighbor bonds forms honeycomb superstructures with the lattice spacing twice longer than the original honeycomb network, as exemplified in Fig.~\ref{mlwf}(c), which leads to new Dirac cones around the midpoints of the $\Gamma$-K lines (zone corners in the folded Brillouin zone).
Thus, in our TMTs, the hidden honeycomb superstructures stemming from the orbital and geometric nature result in the multiple Dirac nodes.

%%%%% Dirac gap by SOC and its potential controllability %%%%%
\subsection{Tunable Dirac gap}
\label{sec:tensile}

\begin{figure}[t]
\includegraphics[width=1.0\columnwidth]{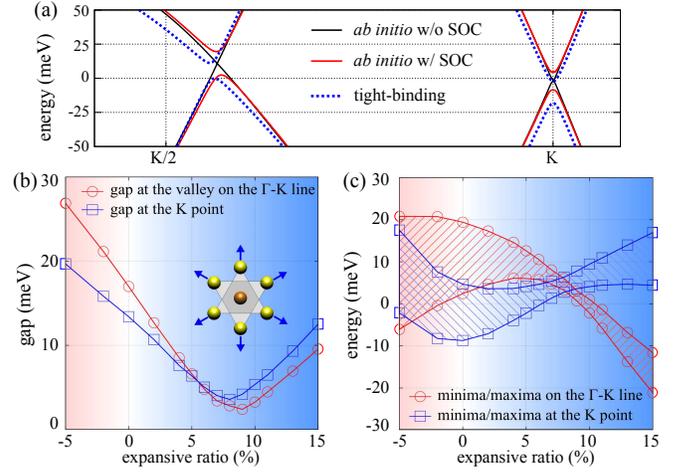}
\caption{
(a)~Enlarged figure of the electronic band structure of PdPS$_{3}$ near the Fermi level along the $\Gamma$-K line.
The black (red) solid lines represent the band dispersions obtained by {\it ab initio} calculations neglecting (including) the SOC.
The blue dotted ones are the dispersions for the tight-binding model with the transfer integrals of MLWFs up to fifth neighboring Pd cations including the effective SOC in Eq.~(\ref{SOC}) with $\tilde{\lambda}=15$~meV. 
(b)~The amplitudes of the Dirac gaps as functions of the expansive ratio of the in-plane lattice constant.
Schematic image of tensile expansion is shown in the inset.
(c)~The tensile-strain dependence of the two valley structures.
}
\label{strain}
\end{figure}

Next, we discuss the effect of SOC.
Although the orbital moment is quenched in the $e_g$ manifold in an ideal octahedral crystal field, the SOC modifies the $e_g$ electronic states through $t_{2g}$-$e_g$ mixing in the presence of a distortion of $MX_6$ octahedra.
Indeed, we find that the Dirac nodes are gapped out by including the relativistic effect in the {\it ab initio} calculations, as shown in Fig.~\ref{strain}(a).
In this monolayer system, the dominant distortion is a trigonal one, which leads to an effective SOC given as~\cite{xiao2011interface} 
\begin{equation} 
H_{\rm SOC}
=
-(\tilde{\lambda}/2)
\sum_{\bm{k}}\sum_{mn}\sum_{\sigma\sigma'}
c^{\dagger}_{\bm{k}m\sigma}(\hat{\tau}_{y})_{mn}(\hat{\sigma}_{z})_{\sigma\sigma'} c_{\bm{k}n\sigma'},
\label{SOC}
\end{equation}
where $c^{\dagger}_{\bm{k} m \sigma}$($c_{\bm{k} m \sigma}$) is the creation (annihilation) operator of an electron for the wave vector $\bm{k}$, orbital $m = d_{3z^{2}-r^{2}}$ or $d_{x^{2}-y^{2}}$, spin $\sigma =\uparrow$ or $\downarrow$, and $\tau_{y}$ ($\sigma_{z}$) is the $y$ ($z$) component of the Pauli matrix for the orbital (spin) space; here, the $xyz$-axes are taken as shown in the inset of Fig.~\ref{mlwf}(a) and the quantization axis of spin is taken along the [111] direction. 
The coupling constant is given as $\tilde{\lambda}=\Delta_{\rm tri}\lambda^{2}/\Delta^{2}$, where  $\Delta$ and $\Delta_{\rm tri}$ are the crystalline electric field from the octahedral ligands and the trigonal distortion, respectively, and $\lambda$ is the coupling constant of the atomic SOC.

Indeed, we confirm that when adding the effective SOC in Eq.~(\ref{SOC}) to the tight-binding model constructed above, its band structures reproduce the gapped Dirac nodes, as shown in Fig.~\ref{strain}(a).
From the comparison, we obtain the rough estimate of the effective SOC $\tilde{\lambda}=15$~meV.
We note that the gapped state is topologically trivial: the $Z_2$ topological invariant~\cite{PhysRevLett.95.146802} becomes zero for all the bands in the tight-binding model.

The result indicates that the Dirac gaps can be controlled through the crystalline symmetry.
Here, we demonstrate it by tensile strain, which has been commonly used for two-dimensional vdW materials~\cite{Lee385,guinea2010energy}.
Starting from the fully-optimized crystalline structure at zero expansive ratio, we extend two in-plane primitive vectors, $\bm{a}$ and $\bm{b}$, while keeping the out-of-plane primitive vector $\bm{c}$.  
We also keep the fractional coordinates of atoms projected onto the $ab$ plane.
Figure~\ref{strain}(b) shows the change in the Dirac gaps.
In the original optimal structure (zero expansive ratio), the octahedra are slightly elongated in the out-of-plane direction. 
While the system is expanded in the in-plane directions, the Dirac gaps decrease and become minimal around $8$-$9$\% expansive ratio, where the trigonal distortion almost vanishes (see Appendix~\ref{app2} for the information on distortions of the octahedra). 
Interestingly, the valley structures of the two massive Dirac cones are shifted individually by the tensile strain, as shown in Fig.~\ref{strain}(c).
These results indicate the flexible tunability of the massive Dirac cones.

%%%%%% Electron correlations and possible topological magnetism %%%%%
\subsection{Topological magnetism induced by electron correlations}
\label{sec:topomag}

\begin{figure}[t]
\includegraphics[width=1.0\columnwidth]{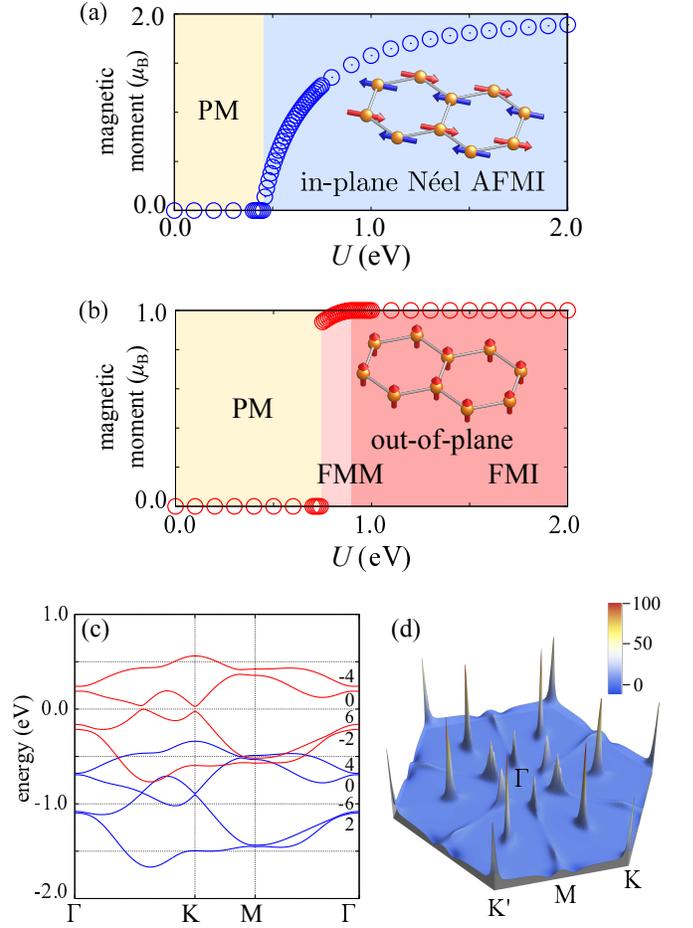}
\caption{
(a) and (b)~Ground-state phase diagrams of the multi-orbital Hubbard model obtained by the mean-field approximation at half filling and 3/4 filling, respectively.
PM, AFMI, FMM, and FMI represent the paramagnetic metal, antiferromagnetic insulator, ferromagnetic metal, and ferromagnetic insulator, respectively. 
The magnitude of magnetic moments is plotted in each magnetic phase.
(c)~Electronic band structure for the FMI at $U=1.5$~eV.
The red and blue lines represent the up and down-spin bands, respectively, and the number on each band indicates the Chern number $C$.
(d)~Wave-number dependence of the Berry curvature of the highest-occupied band with $C=6$ in the first Brillouin zone. 
}
\label{mf}
\end{figure}

Let us discuss the effect of electron correlations in $d$ electrons on the multiple-Dirac semimetals.
Although the magnetism was studied for monolayer TMTs by {\it ab initio} calculations~\cite{PhysRevB.91.235425,PhysRevB.94.184428}, the previous works focused on the $3d$ compounds in which the SOC is irrelevant. 
We here investigate the synergetic effect of electron correlations and the SOC, both of which can be relevant in $4d$ and $5d$ compounds, using an effective multi-orbital Hubbard model and the mean-filed approximation. 
The multi-orbital Hubbard model consists of the tight-binding model with the transfer integrals in Table~\ref{hwr}, the effective SOC in Eq.~(\ref{SOC}) with $\tilde{\lambda}=$15 meV, and the onsite Coulomb interactions given by 
\begin{equation}
H_{\rm int}
=
\frac{1}{2} \sum_{mnm'n'} U_{mnm'n'}
\sum_{i}
\sum_{\sigma \sigma'}
c^{\dagger}_{i m\sigma} c^{\dagger}_{i n\sigma'} c_{i n'\sigma'} c_{im'\sigma},
\label{onsite}
\end{equation}
where $c^{\dagger}_{i m \sigma}$($c_{i m \sigma}$) is the Fourier transform of $c^{\dagger}_{\bm{k} m \sigma}$($c_{\bm{k} m \sigma}$).
Assuming the rotational symmetry of the Coulomb interaction, we set $U_{mmmm} = U$,  $U_{mnmn} = U-2J$, and $U_{mnnm}=U_{mmnn}=J$ ($m\neq n$), where $U$ is the intraorbital Coulomb interaction and $J$ is the Hund's coupling, respectively; we take $J/U=0.2$ in the following calculations.
In the mean-field calculation, we adopt the standard Hartree-Fock approximation to decouple the onsite interaction terms in Eq.~(\ref{onsite}). 
We take into account charge, spin, and orbital orders with the ordering vector $\bm{Q}=(0, 0)$ or $(\pi,\pi)$ on the honeycomb lattice and approximate the integration in the first Brillouin zone by the summation over 128$\times$128 $\bm{k}$ points and determine the mean fields consistently within a precision of less than 10$^{-6}$.

We focus on two commensurate fillings, half filling (two $e_g$ electrons per $M$) and 3/4 filling (three $e_g$ electrons per $M$); the former corresponds to the situation discussed above, while the latter a chemical substitution of $M$ by, e.g., Ag or Cd. 
We note that such substitutions were reports for the bulk form of TMTs~\cite{BREC19863}.
Figures \ref{mf}(a) and \ref{mf}(b) show the ground-state phase diagrams and the magnetic moments, obtained by the mean-field analysis in the range of Coulomb interactions including realistic values for 4$d$ and 5$d$ transition metal  compounds~\cite{0953-8984-29-26-263001}.

At half filling [Fig.~\ref{mf}(a)], while increasing the electron interactions, the system exhibits a continuous phase transition from the paramagnetic Dirac semimetal to a N{\'e}el-type antiferromagnetic insulator (AFMI) with in-plane magnetic moments.
For comparison, we also performed the {\it ab initio} calculations in the GGA scheme with allowing magnetic solutions (see Appendix~\ref{app3} for the details).
We find that, similar to the mean-field results, the lowest-energy state changes from the paramagnetic Dirac semimetal to AFMI while changing from weakly correlated $M$=Pt to strongly correlated $M$=Ni; the $M$=Pd case is close to the boarder.
We note that, in general, the {\it ab initio} calculation tends to underestimate 
the correlation effects while the mean-field approximation tends to overestimate. 
From these considerations, we conclude that PdPS$_3$ might be in the AFM phase, while PtPS$_3$ the multiple-Dirac semimetal (see also the discussion in Appendix~\ref{app3}). 
As PdPS$_3$ appears to locate close to the border, it might also be possible to transform it to the multiple-Dirac semimetal by tuning the bandwidth by the substitution of S by Se (see Appendix~\ref{app1}) or by tensile strain.

On the other hand, at 3/4 filling, the system shows a discontinuous phase transition from the paramagnetic metal to a ferromagnetic metal (FMM), and to a ferromagnetic insulator (FMI) with out-of-plane magnetic moments, as shown in Fig.~\ref{mf}(b).
For comparison, we also performed the GGA calculation for AgPS$_3$, which reproduces the 3/4-filled state, using the crystalline structure of PdPS$_3$~\cite{ZAAC:ZAAC19733960305}.
We found that the lowest-energy state is the FM state with a small magnetic moment, 0.04 $\mu_{\rm B}$, which might also reflect the general tendency of GGA calculations to underestimate electron interactions.

Interestingly, we find that the ferromagnetic states at 3/4 filling acquires nontrivial topological nature.
Figure \ref{mf}(c) shows the band structure of the FMI at $U=1.5$~eV.
The bands are split by the exchange field into the up-spin (red) and down-spin (blue) ones, and the lower six are occupied at 3/4 filling.
Note that as the mean-field Hamiltonian in the ferromagnetic state conserves the spin $z$ component, we can separate the mean-field Hamiltonian into up-spin and down-spin sectors and distinguish the spin state of each electronic band in Fig.~\ref{mf}(c).
We calculate the Berry curvature and the Chern number for each band of the mean-field solution by using the standard Kubo formula~\cite{PhysRevLett.49.405}.
Summing the Chern number of the occupied bands [see Fig.~\ref{mf}(c)], we find that the FMI is a topologically-nontrivial ferromagnet with rather high Chern number $C=4$.
Figure~\ref{mf}(d) displays the wave-number dependence of the Berry curvature of the highest-occupied band with $C=6$. 
The Berry curvatures shows spikes at the K and K' points and around the midpoints of the $\Gamma$-K lines. 
These anomalous contributions can be traced back to the Dirac cones in the original semimetallic state.
Thus, our results suggest that the multiple-Dirac semimetal can be turned into an unconventional topological ferromagnet with high Chern number by electron correlations and carrier doping.

%%%%% Discussion %%%%%
\section{Summary and concluding remarks}
\label{sum}
To summarize, we have theoretically uncovered two potential electronic properties of TMTs with 4$d$ and 5$d$ transition metals in the monolayer form.
One is the highly-tunable multiple Dirac cones.
This will bring about new transport phenomena, such as the unconventional Hall responses and the multiple valley operations.  
The other is the topological ferromagnetism with high Chern number driven by electron correlations and chemical doping.
This will provide new candidates for quantized anomalous Hall insulators, whose multiple chiral edge modes might be used for a thin-film transmitter with high efficiency. 
We believe that the two features will stimulate further material exploration in 4$d$ and 5$d$ TMTs for delivering missing pieces in material science of atomically-thin films and the heterostructures.  

The recipe for multiple Dirac cones found here is generic and simple: transition metal cations with $e_g$ orbitals and octahedral ligands forming edge-sharing honeycomb structure with extended $p$ orbitals.
As such a crystalline and electronic structure is widely seen in layered transition metal compounds, e.g., transition metal chalcogenides and halides~\cite{Hulliger1976}, it is of great interest to search other candidates of multiple-Dirac semimetals.
Another intriguing issue would be the effect of electron correlations on the Dirac cones, which has been intensively studied in graphene and related compounds. 
The additional valley degrees of freedom in our multiple Dirac nodes  may give rise to richer physics between the Dirac semimetal and the Mott insulator.
Therefore, we believe that our findings stimulate the further exploration of TMTs and their relatives, which would pave a way to post-graphene nanotechnology.

%%%%% Acknowledgement %%%%%
\begin{acknowledgments}
Y.S. is supported by the Japan Society for the Promotion of Science through a research fellowship for young scientists and the Program for Leading Graduate Schools (MERIT).
This research was supported by Grants-in-Aid for Scientific Research under Grants No. JP15K05176.
The crystal structures and MLWFs are visualized by using VESTA 3~\cite{Momma:db5098}.
\end{acknowledgments}

\begin{appendix}
%%%%% Band structures %%%%%%
\section{{\it Ab initio} results for monolayer $M$P$X_{3}$ in the paramagnetic state}
\label{app1}

\begin{figure}[t]
\centering
\includegraphics[width=1.0\columnwidth]{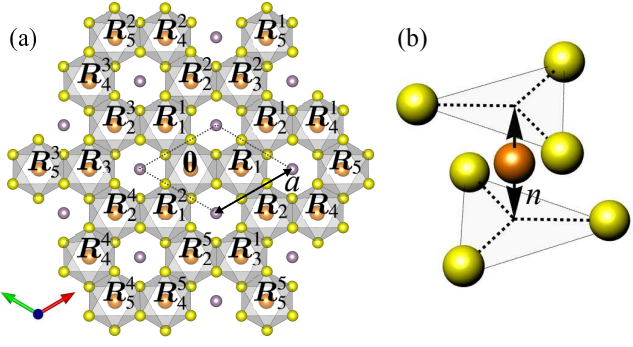}
\caption{
(a)~Schematic picture of optimized honeycomb-monolayer structure.
The dotted lines and the solid arrow indicate the unit cell and the lattice constant $a$ (see Table~\ref{lattice}).
Atomic positions used in the calculation of transfer integrals in Table~\ref{all} are also denoted.
(b)~Schematic picture of transition metal cation $M$ sandwiched by $X_{3}$ triangles.
We define the layer thickness $n$ as the distance between centers of the upper and lower $X_{3}$ triangles in the unit cell (see Table~\ref{lattice}).
}
\label{unit}
\end{figure}

\begin{figure}[!h]
\centering
\includegraphics[width=0.8\columnwidth]{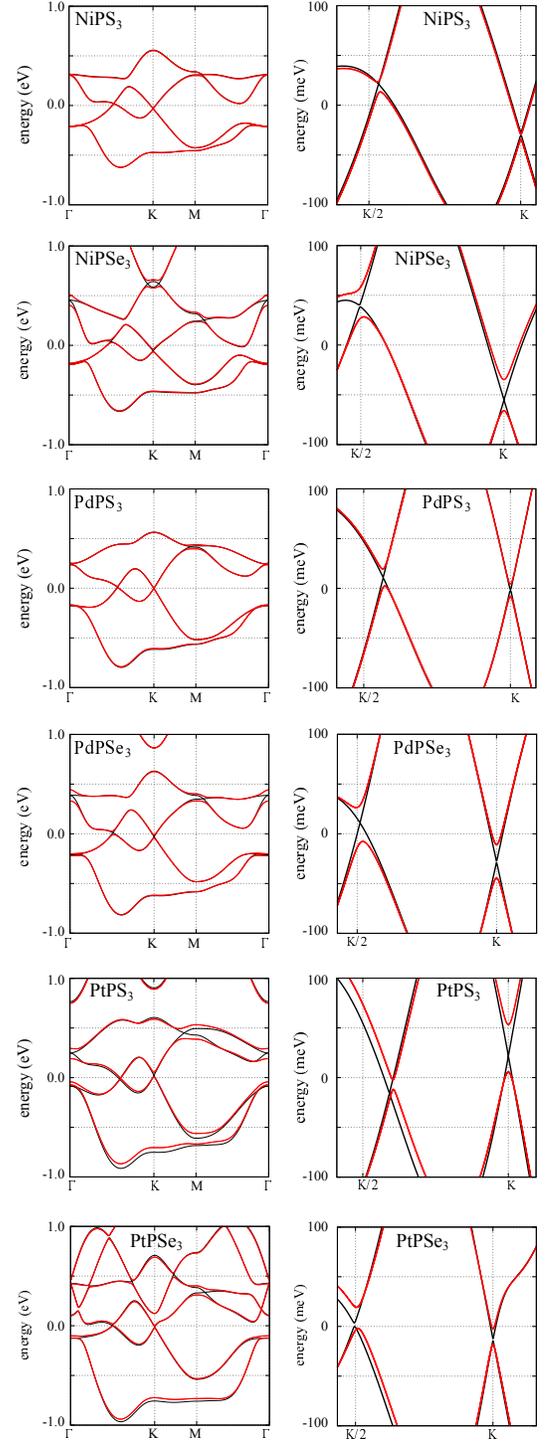}
\caption{
Electronic band structures of monolayer $M$P$X_{3}$ with $M$=Ni, Pd, and Pt, and $X$= S and Se.
Left panels show the entire range of the $e_g$ bands.
Right panels show the enlarged figures near the Fermi level along the $\Gamma$-K line.
The black (red) solid lines represent the band dispersions obtained by {\it ab initio} calculations neglecting (including) the SOC.
}
\label{band}
\end{figure}

In this Appendix, we show the details of {\it ab initio} results for $M$P$X_{3}$ ($M$=Ni, Pd, and Pt, and $X$=S and Se).
We confirm that the honeycomb monolayer form (Fig.~\ref{unit}) is structurally stable for all $M$P$X_3$, though the lattice constant $a$ is different; see Table~\ref{lattice}.
We note that the optimized structures are elongated along the out-of-plane direction compared to the ideal octahedra: in all cases, the ratio $a/n$ (see Fig.~\ref{unit}) is smaller than the ideal value $3/\sqrt{2} \sim 2.12$.

Figure \ref{band} shows the electronic band structures of monolayer $M$P$X_{3}$ obtained by {\it ab initio} calculations with and without the SOC.
In all cases, the band structures show multiple Dirac dispersions at the middle points on the $\Gamma$-K lines as well as the K and K' points near the Fermi level, which are gapped out by the SOC.
This indicates that the multiple Dirac cones originating from the orbital and geometric nature discussed in Sec.~\ref{sec:multipleDirac} are ubiquitous in the monolayer TMTs.
We note that similar band dispersions were already seen in the previous {\it ab initio} studies~\cite{PhysRevB.94.184428} although less attention has been paid.
We also note that the previous experimental work reported insulating behavior even above the N\'{e}el temperature for NiPS$_{3}$ in the bulk form~\cite{ic50197a018}, while our result for the monolayer form is semimetallic in the paramagnetic state.
The discrepancy would be ascribed to the difference between bulk and monolayer or the importance of electron interactions in 3$d$ electrons, which is in general underestimated in {\it ab initio} calculations.

\begin{table}[b]
\begin{center}
\begin{ruledtabular}
\begin{tabular}{l|ccc|ccc}
		&NiPS$_{3}$ 	&PdPS$_{3}$	&PtPS$_{3}$ 	&NiPSe$_{3}$ 	&PdPSe$_{3}$ &PtPSe$_{3}$\\
\hline
$a$\,(\AA)	&$5.82$		&$6.02$		&$6.07$		&$6.17$		&$6.34$		&$6.42$\\
$n$\,(\AA)	&$2.98$		&$3.17$		&$3.07$		&$3.07$		&$3.27$		&$3.12$\\
$a/n$	&$1.95$		&$1.90$		&$1.97$		&$2.01$		&$1.94$		&$2.05$\\
\end{tabular}
\end{ruledtabular}
\end{center}
\caption{
The lattice constant and the layer thickness of optimized structures of $M$P$X_{3}$.
Schematic pictures of  $a$ and $n$ are illustrated in Fig.~\ref{unit}.
In the honeycomb network of the edge-sharing ideal octahedra, the ratio of the parameters should be $a/n =3/\sqrt{2} \sim 2.12$.
}
\label{lattice}
\end{table}

In addition, we provide the detailed information on the transfer integrals estimated by MLWFs for PdPS$_3$.
While the representative values are shown in Table~\ref{mlwf}, we enlist all the values in Table~\ref{all} up to fifth neighbors. 
The spatial positions of the neighbors are illustrated in Fig.~\ref{unit}(a).
As discussed in Sec.~\ref{sec:multipleDirac}, the transfer integrals for third neighbors are the most dominant among them on average.
Note that the transfer integrals for different directions in the same distance are related with each other via the point-group operations; for instance, when the $MX_6$ octahedra have $D_{3h}$ symmetry, the hopping matrices should follow $\hat{h}(\hat{C_3} \bm{R}) =\hat{R}_{C_3} \hat{h}(\bm{R}) \hat{R}^{-1}_{C_3}$, where $(\hat{h}(\bm{R}))_{mn} = \bra{m,\bm{0}} H \ket{n,\bm{R}}$ and $\hat{R}_{C_3}$ is the threefold-rotation operator on the $e_g$-orbital basis.
Indeed, Table~\ref{all}  indicates that such relations hold approximately for the optimized structure in the {\it ab initio} calculation. 

\begin{table}
\begin{ruledtabular}
\begin{tabular}{l|ccc|cccccc}
$(m,n)$	&$\bm{R}_{1}$	&$\bm{R}^{1}_{1}$	&$\bm{R}^{2}_{1}$	&$\bm{R}_{2}$	&$\bm{R}^{1}_{2}$	&$\bm{R}^{2}_{2}$	&$\bm{R}^{3}_{2}$	&$\bm{R}^{4}_{2}$	&$\bm{R}^{5}_{2}$\\
\hline
$(3z^{2}-r^{2},3z^{2}-r^{2})$	&$-87$	&$-72$	&$-72$	&$-9$	&$-9$	&$26$	&$-9$	&$-9$	&$26$\\
$(3z^{2}-r^{2},x^{2}-y^{2})$	&$0$		&$9$		&$-9$	&$22$	&$-22$	&$2$		&$18$	&$-18$	&$-2$\\
$(x^{2}-y^{2},3z^{2}-r^{2})$	&$0$		&$9$		&$-9$	&$18$	&$-18$	&$-2$	&$22$	&$-22$	&$2$	\\
$(x^{2}-y^{2},x^{2}-y^{2})$		&$-70$	&$-82$	&$-82$	&$14$	&$14$	&$-21$	&$14$	&$14$	&$-21$\\
\end{tabular}
\begin{tabular}{l|ccc|cccccc}
$(m,n)$					&$\bm{R}_{3}$	&$\bm{R}^{1}_{3}$	&$\bm{R}^{2}_{3}$	&$\bm{R}_{4}$	&$\bm{R}^{1}_{4}$	&$\bm{R}^{2}_{4}$	&$\bm{R}^{3}_{4}$	&$\bm{R}^{4}_{4}$	&$\bm{R}^{5}_{4}$\\
\hline
$(3z^{2}-r^{2},3z^{2}-r^{2})$	&$-38$		&$219$			&$219$			&$4$			&$4$				&$-1$			&$13$			&$13$			&$-1$\\
$(3z^{2}-r^{2},x^{2}-y^{2})$	&$0$			&$147$			&$-147$			&$-8$		&$8$				&$5$				&$-3$			&$3$				&$-5$\\
$(x^{2}-y^{2},3z^{2}-r^{2})$	&$0$			&$147$			&$-147$			&$-8$		&$8$				&$5$				&$-3$			&$3$				&$-5$\\
$(x^{2}-y^{2},x^{2}-y^{2})$		&$304$		&$48$			&$48$			&$7$			&$7$				&$12$			&$-2$			&$-2$			&$12$\\
\end{tabular}
\begin{tabular}{l|cccccc}
$(m,n)$					&$\bm{R}_{5}$	&$\bm{R}^{1}_{5}$	&$\bm{R}^{2}_{5}$	&$\bm{R}^{3}_{5}$	&$\bm{R}^{4}_{5}$	&$\bm{R}^{5}_{5}$\\
\hline
$(3z^{2}-r^{2},3z^{2}-r^{2})$	&$-12$		&$19$			&$19$			&$-12$			&$19$			&$19$\\
$(3z^{2}-r^{2},x^{2}-y^{2})$	&$0$			&$-18$			&$18$			&$0$				&$-18$			&$18$\\
$(x^{2}-y^{2},3z^{2}-r^{2})$	&$0$			&$-18$			&$18$			&$0$				&$-18$			&$18$\\
$(x^{2}-y^{2},x^{2}-y^{2})$		&$30$		&$-1$			&$-1$			&$30$			&$-1$			&$-1$\\
\end{tabular}

\end{ruledtabular}
\caption{
Transfer integrals between MLWFs.
Each value in the table means $\bra{m,\bm{0}} H \ket{n,\bm{r}}$, where $H$ is the Hamiltonian of the system and $\ket{m,\bm{r}}$ is the $d_{m}$-like MLWF at site $\bm{r}$ ($m$ = $3z^{2}-r^{2}$ or $x^{2}-y^{2}$).
See Fig.~\ref{unit}(a) for the spatial positions of $\bm{r}$.
The unit of transfer integrals is in meV.
}
\label{all}
\end{table}

%%%%% Elongation %%%%%%
\section{Distortion of octahedra by the tensile strain}
\label{app2}
\begin{figure}[t]
\centering
\includegraphics[width=0.9\columnwidth]{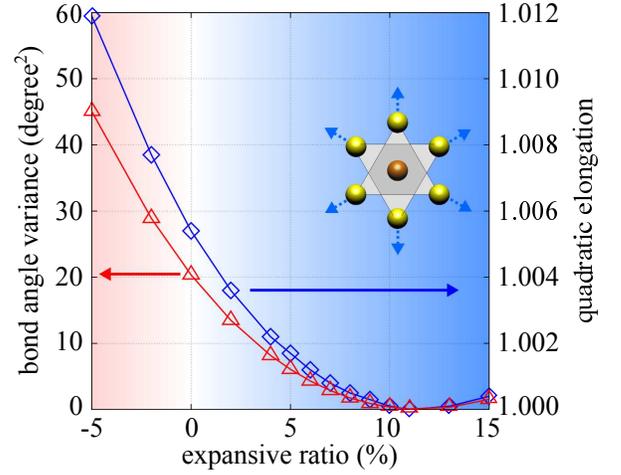}
\caption{
The bond angle variance and the quadratic elongation of PdS$_{6}$ octahedra as functions of the expansive ratio of the in-plane lattice constant obtained by the {\it ab initio} calculations.
}
\label{elo}
\end{figure}

In this Appendix, we show how PdS$_6$ octahedra are distorted for the tensile strain discussed in Sec.~\ref{sec:tensile}.
We adopt two conventional measures for the distortion~\cite{robinson1971quadratic}: one is the bond angle variance $\sum^{12}_{i=1}(\theta_i - 90^{\circ})^2/11$, where $\theta_i$ is the angle of a S-Pd-S bond for neighboring S, and the other is the quadratic elongation $\sum^{6}_{i=1}(l_{i}/l_{0})^2/6$, where $l_{i}$ is the length of a Pd-S bond and $l_0$ is the bond length of an ideal octahedron with the same volume.
Figure~\ref{elo} shows the the bond angle variance and the quadratic elongation of PdS$_6$ octahedra as functions of the expansive ratio obtained by our {\it ab initio} calculations.
The result indicates that PdS$_6$ octahedra is originally elongated along the $c$ axis at the zero expansive ratio (see also Table~\ref{lattice}) and approaches to ideal octahedra while increasing the expansive ratio; the distortion is minimized around 10\%.
This is consistent with the behavior of the Dirac gaps plotted in Fig.~3(b), which is predominantly opened by the trigonal distortion of the octahedra. 

%%%%% Magnetism %%%%%%
\section{Total energy comparison by GGA calculation}
\label{app3}

\begin{figure}[b]
\centering
\includegraphics[width=0.8\columnwidth]{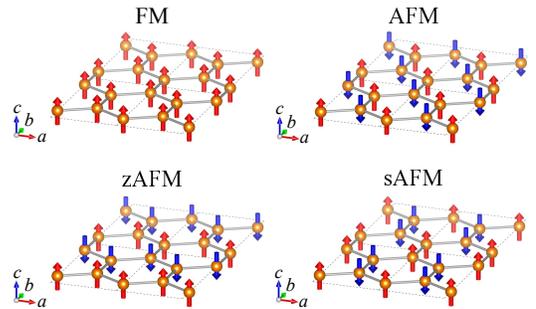}
\caption{
Schematic pictures of magnetic configurations considered in the GGA calculations.
The rectangular unit cells are indicated by the dotted lines.
}
\label{rec}
\end{figure}

In this Appendix, we show the {\it ab initio} results while allowing magnetic solutions. 
Using GGA calculations without the SOC under the full structural optimization, we compute the total energy of paramagnetic (PM), ferromagnetic (FM), N\'{e}el antiferromagnetic (AFM), zigzag antiferromagnetic (zAFM), and stripy antiferromagnetic (sAFM) states.
We adopted a rectangular unit cell (see Fig.~\ref{rec}) and $16\times16\times1$ $\bm{k}$-point mesh.

Table~\ref{gga} shows the total energy and the magnetic moment in each state.
The lowest energy solution is for the zAFM for the $3d$ electron system with $M$=Ni, while it is PM for the $5d$ $M$=Pt.
We note that the zAFM was indeed found for the bulk NiPS$_3$~\cite{BREC19863}.
In the case of the $4d$ $M$=Pd, the lowest-energy state is the AFM, while the obtained solutions are almost degenerate in energy among PM, AFM, and zAFM.
The results indicate the trend from strongly correlated $3d$ to weakly correlated $5d$; the $4d$ case is in the boarder between the magnetic insulator and the Dirac paramagnetic semimetal.
Thus, PtPS$_3$ would be a prime candidate for the multiple Dirac semimetal rather than PdPS$_3$.
In addition, from the comparison of the magnetic moments between the GGA calculation for PdPS$_3$ and the mean-field analysis at half filling [see Fig. 4(a)], the corresponding $U$ value in the mean-field analysis is about 0.5~eV, which appears to be weak for 4$d$ transition compounds.
This may be ascribed to the generic tendency that the GGA calculation underestimates the electron interactions of $d$ orbitals.

\begin{table}[!h]
\begin{center}
\begin{ruledtabular}
\begin{tabular}{l|ccccc}
			&PM 	&FM				&AFM		&zAFM		&sAFM\\
\hline
NiPS$_{3}	$	&$914$	&$444\,\,(1.24)$	&$41\,\,(1.13)$	&$0\,\,(1.12)$	&$491\,\,(1.24)$\\
PdPS$_{3}$	&$5$		&--				&$0\,\,(0.26)$	&$1\,\,(0.27)$	&--\\
PtPS$_{3}$	&$0$		&--				&--			&--			&--\\
\end{tabular}
\end{ruledtabular}
\end{center}
\caption{
The total energy of each electronic state obtained by GGA calculations.
The lowest energy of all electronic states is set to be zero for each compound.
The blanks indicate that the corresponding state is not obtained as a stable solution.
The unit of energy is in meV per rectangular unit cell including four formula units (see Fig.~\ref{rec}).
For the magnetic solutions, we denote the value of the magnetic moment per transition metal cation in unit of Bohr magneton $\mu_{\rm B}$.
}
\label{gga}
\end{table}
\end{appendix}

%%%%% References %%%%%

\end{document}